\documentstyle[twoside,epsf]{article}

\catcode`\@=11
\long\def\@makefntext#1{
\protect\noindent \hbox to 3.2pt {\hskip-.9pt
$^{{\eightrm\@thefnmark}}$\hfil}#1\hfill}       

\def\thefootnote{\fnsymbol{footnote}}
\def\@makefnmark{\hbox to 0pt{$^{\@thefnmark}$\hss}}    

\def\ps@myheadings{\let\@mkboth\@gobbletwo
\def\@oddhead{\hbox{}
\rightmark\hfil\eightrm\thepage}
\def\@oddfoot{}\def\@evenhead{\eightrm\thepage\hfil
\leftmark\hbox{}}\def\@evenfoot{}
\def\sectionmark##1{}\def\subsectionmark##1{}}

\oddsidemargin=\evensidemargin
\renewcommand{\thefootnote}{\fnsymbol{footnote}}

\newcounter{sectionc}\newcounter{subsectionc}\newcounter{subsubsectionc}
\renewcommand{\section}[1] {\vspace{12pt}\addtocounter{sectionc}{1}
\setcounter{subsectionc}{0}\setcounter{subsubsectionc}{0}\noindent
    {\tenbf\thesectionc. #1}\par\vspace{5pt}}
\renewcommand{\subsection}[1] {\vspace{12pt}\addtocounter{subsectionc}{1}
\setcounter{subsubsectionc}{0}\noindent
{\bf\thesectionc.\thesubsectionc. {\kern1pt \bfit #1}}\par\vspace{5pt}}
\renewcommand{\subsubsection}[1] {\vspace{12pt}\addtocounter{subsubsectionc}{1}
    \noindent{\tenrm\thesectionc.\thesubsectionc.\thesubsubsectionc.
    {\kern1pt \tenit #1}}\par\vspace{5pt}}
\newcommand{\nonumsection}[1] {\vspace{12pt}\noindent{\tenbf #1}
    \par\vspace{5pt}}

\newcounter{appendixc}
\newcounter{subappendixc}[appendixc]
\newcounter{subsubappendixc}[subappendixc]
\renewcommand{\thesubappendixc}{\Alph{appendixc}.\arabic{subappendixc}}
\renewcommand{\thesubsubappendixc}
    {\Alph{appendixc}.\arabic{subappendixc}.\arabic{subsubappendixc}}

\renewcommand{\appendix}[1] {\vspace{12pt}
        \refstepcounter{appendixc}
        \setcounter{figure}{0}
        \setcounter{table}{0}
        \setcounter{lemma}{0}
        \setcounter{theorem}{0}
        \setcounter{corollary}{0}
        \setcounter{definition}{0}
        \setcounter{equation}{0}
        \renewcommand{\thefigure}{\Alph{appendixc}.\arabic{figure}}
        \renewcommand{\thetable}{\Alph{appendixc}.\arabic{table}}
        \renewcommand{\theappendixc}{\Alph{appendixc}}
        \renewcommand{\thelemma}{\Alph{appendixc}.\arabic{lemma}}
        \renewcommand{\thetheorem}{\Alph{appendixc}.\arabic{theorem}}
        \renewcommand{\thedefinition}{\Alph{appendixc}.\arabic{definition}}
        \renewcommand{\thecorollary}{\Alph{appendixc}.\arabic{corollary}}
        \renewcommand{\theequation}{\Alph{appendixc}.\arabic{equation}}
        \noindent{\tenbf Appendix \theappendixc #1}\par\vspace{5pt}}
\newcommand{\subappendix}[1] {\vspace{12pt}
        \refstepcounter{subappendixc}
        \noindent{\bf Appendix \thesubappendixc. {\kern1pt \bfit #1}}
    \par\vspace{5pt}}
\newcommand{\subsubappendix}[1] {\vspace{12pt}
        \refstepcounter{subsubappendixc}
        \noindent{\rm Appendix \thesubsubappendixc. {\kern1pt \tenit #1}}
    \par\vspace{5pt}}

\newcommand{\textlineskip}{\baselineskip=13pt}
\newcommand{\smalllineskip}{\baselineskip=10pt}

\newcommand{\copyrightheading}[1]
    {\vspace*{-2.5cm}\smalllineskip{\flushleft
    {\footnotesize Quantum Information and Computation, Vol.~1, No.~0 (2012) 000--000 #1}\\
    {\footnotesize \copyright\kern2pt Rinton Press}\\
     }}

\newcommand{\publisher}[2]{{\begin{center}\footnotesize\smalllineskip
    Received #1\\
    Revised #2
    \end{center}
    }}

\def\abstracts#1#2#3{{
    \centering{\begin{minipage}{4.5in}\footnotesize\baselineskip=10pt
    \parindent=0pt #1\par
    \parindent=15pt #2\par
    \parindent=15pt #3
    \end{minipage}}\par}}

\def\keywords#1{{
    \centering{\begin{minipage}{4.5in}\footnotesize\baselineskip=10pt
    {\footnotesize\it Keywords}\/: #1
     \end{minipage}}\par}}
\def\communicate#1{{
    \centering{\begin{minipage}{4.5in}\footnotesize\baselineskip=10pt
    {\footnotesize\it Communicated by}\/: #1
     \end{minipage}}\par}}

\renewenvironment{thebibliography}[1]
        {\frenchspacing
     \ninerm\baselineskip=11pt
         \begin{list}{\arabic{enumi}.}
        {\usecounter{enumi}\setlength{\parsep}{0pt}
     \setlength{\leftmargin 12.7pt}{\rightmargin 0pt}
         \setlength{\itemsep}{0pt} \settowidth
    {\labelwidth}{#1.}\sloppy}}{\end{list}}

\newcounter{itemlistc}
\newcounter{romanlistc}
\newcounter{alphlistc}
\newcounter{arabiclistc}

\newcommand{\fcaption}[1]{
        \refstepcounter{figure}
        \setbox\@tempboxa = \hbox{\footnotesize Fig.~\thefigure. #1}
        \ifdim \wd\@tempboxa > 5in
           {\begin{center}
        \parbox{5in}{\footnotesize\smalllineskip Fig.~\thefigure. #1}
            \end{center}}
        \else
             {\begin{center}
             {\footnotesize Fig.~\thefigure. #1}
              \end{center}}
        \fi}

\newcommand{\tcaption}[1]{
        \refstepcounter{table}
        \setbox\@tempboxa = \hbox{\footnotesize Table~\thetable. #1}
        \ifdim \wd\@tempboxa > 5in
           {\begin{center}
        \parbox{5in}{\footnotesize\smalllineskip Table~\thetable. #1}
            \end{center}}
        \else
             {\begin{center}
             {\footnotesize Table~\thetable. #1}
              \end{center}}
        \fi}

\def\pmb#1{\setbox0=\hbox{#1}
    \kern-.025em\copy0\kern-\wd0
    \kern.05em\copy0\kern-\wd0
    \kern-.025em\raise.0433em\box0}

\def\fnt#1#2{\footnotetext{\kern-.3em
    {$^{\mbox{\scriptsize #1}}$}{#2}}}

\def\fpage#1{\begingroup
\voffset=.3in
\thispagestyle{empty}\begin{table}[b]\centerline{\footnotesize #1}
    \end{table}\endgroup}

\def\runninghead#1#2{\pagestyle{myheadings}
\markboth{{\protect\footnotesize\it{\quad #1}}\hfill}
{\hfill{\protect\footnotesize\it{#2\quad}}}}
\font\tenrm=cmr10
\font\tenit=cmti10
\font\tenbf=cmbx10
\font\bfit=cmbxti10 at 10pt
\font\ninerm=cmr9

\font\eightrm=cmr8

\def\FigName{figure}%
\newbox\captionbox
\long\def\@makecaption#1#2{%
  \ifx\FigName\@captype
    \vskip\abovecaptionskip
    \setbox\tempbox\hbox{{\figurecaptionfont #1\hskip1em #2}}
    \ifdim\wd\tempbox< 28pc
    \centerline{\box\tempbox}
    \else
    {\figurecaptionfont #1\hskip1em #2\par}
\fi\else
    \setbox\tempbox\hbox{{\tablecaptionfont #1\hskip1em #2}}
    \ifdim\wd\tempbox< 28pc
    \centerline{\box\tempbox}
    \else
    {\tablecaptionfont #1\hskip1em #2\par}%
    \fi
 \vskip\belowcaptionskip
 \fi}
\InputIfFileExists{psfig.sty}
{\typeout{^^Jpsfig.sty inputed...ok}}{\typeout{^^JWarning: psfig.sty could be be found.^^J}}
\InputIfFileExists{epsfsafe.tex}
{\typeout{^^Jepsfsafe.tex inputed...ok}}
            {\typeout{^^JWarning: epsfsafe.tex could not be found.^^J}}
\InputIfFileExists{epsfig.sty}
{\typeout{^^Jepsfig.sty inputed...ok}}{\typeout{^^JWarning: epsfig.sty could not be found.^^J}}
\InputIfFileExists{epsf.sty}
{\typeout{^^Jepsf.sty inputed...ok}}{\typeout{^^JWarning: epsf.sty could not be found.^^J}}%
\def\fps@figure{tbp}
\def\ftype@figure{1}
\def\ext@figure{lof}
\def\fnum@figure{Fig.\ \thefigure}
\def\qed{\hbox{${\vcenter{\vbox{              
   \hrule height 0.4pt\hbox{\vrule width 0.4pt height 6pt
   \kern5pt\vrule width 0.4pt}\hrule height 0.4pt}}}$}}

\renewcommand{\thefootnote}{\fnsymbol{footnote}}  

\begin{document}
\setlength{\textheight}{8.0truein}    

\runninghead{Title   $\ldots$}
            {Author(s) $\ldots$}

\normalsize\textlineskip
\thispagestyle{empty}
\setcounter{page}{1}

\copyrightheading{} 

\vspace*{0.88truein}

\fpage{1}
\centerline{\bf
A NOTE ON LOCALLY UNEXTENDIBLE NON-MAXIMALLY } \vspace*{0.035truein}
\centerline{\bf ENTANGLED BASIS}
\vspace*{0.035truein}
\centerline{\bf --- Comment on ``Quant. Inf. Comput. 12, 0271(2012)"}
\vspace*{0.37truein}
\centerline{\footnotesize
BIN CHEN \footnote{Email:chenbin5134@163.com} \ \ \ HALQEM NIZAMIDIN \footnote{Email:nhalqem@gmail.com}
\ \ \ SHAO-MING FEI \footnote{Email:feishm@mail.cnu.edu.cn}}
\vspace*{0.015truein}
\centerline{\it School of Mathematical Sciences, Capital Normal University}
\baselineskip=10pt
\centerline{\it Beijing, 100048, China}
\vspace*{0.225truein}
\publisher{(received date)}{(revised date)}

\vspace*{0.21truein}
\abstracts{
We study the locally unextendible non-maximally entangled basis
(LUNMEB) in $H^{d}\bigotimes H^{d}$. We point out that there exists
an error in the proof of the main result of LUNMEB [Quant. Inf. Comput. 12, 0271(2012)],
which claims that there are
at most $d$ orthogonal vectors in a LUNMEB, constructed from a given non-maximally
entangled state. We show that both the proof and the main result
are not correct in general. We present a
counter example for $d=4$, in which five orthogonal vectors from a specific non-maximally entangled
state are constructed. Besides, we completely solve the problem of LUNMEB for the case of $d=2$.}{}{}

\vspace*{10pt}
\vspace*{3pt}
\communicate{to be filled by the Editorial}

\vspace*{1pt}\textlineskip  
\section{Introduction}          
\vspace*{-0.5pt}
\noindent
The unextendible product basis (UPB) has been extensively investigated.
Considerable elegant results have been obtained with interesting applications to
the theory of quantum information [1,2,3]. Recently S. Bravyi and J. A.
Smolin generalized the notion of the UPB to unextendible
entangled basis. They studied the special case -- unextendible
maximally entangled basis (UMEB) in $H^{d}\bigotimes H^{d}$ for
$d=2,3,4$ [4]. After that I. Chakrabarty, P. Agrawal and A.K. Paty
introduced the concept of locally unextendible non-maximally
entangled basis (LUNMEB) in $H^{d}\bigotimes H^{d}$ [5]: A set of
states $\{|\phi_{a}\rangle\in H^{d}\bigotimes H^{d},~
a=1,2,\cdots,n\}$ is called LUNMEB if and only if

(i) all states $|\phi_{a}\rangle$ are non-maximally entangled;

(ii) $\langle\phi_{a}|\phi_{b}\rangle=\delta_{a,b}$;

(iii) $\forall~ a,b$, there exits a unitary operator $U_{ab}$ such that
$|\phi_{a}\rangle=(U_{ab}\bigotimes I)|\phi_{b}\rangle$;

(iv) if $\langle\Phi|\phi_{a}\rangle=0$ for all $a$, then there is no
unitary operator $V$ such that $(V\bigotimes
I)|\phi_{a}\rangle=|\Phi\rangle$ for some $a$.

As the local unitary transformations do not change the degree of entanglement,
all the basic vectors $|\phi_{a}\rangle$ in LUNMEB have the same
entanglement. The main theorem in [5] claims that starting from a general non-maximally entangled state,
one can get at most $d$ orthogonal vectors to form a LUNMEB by applying a set of
specific local unitary transformations. However, the proof of this
theorem has a flaw, and the main result is also not correct in general.
We can give a counter example to show the theorem is wrong for $d=4$.

In addition, we study thoroughly the case of $d=2$. In [5] special
unitary operators are used to show that there are no more than two
orthogonal vectors in LUNMEB for $d=2$. We will show that
for $d=2$, any unitary transformation applying to one party of a given
non-maximally entangled state can ensure that there does not exist
the third vector in constructing a LUNMEB. Thus we give a complete
result of LUNMEB for $H^{2}\bigotimes H^{2}$.

\section{Counter example in $H^{4}\bigotimes H^{4}$}
\noindent We first review the proof of the main theorem in [5] which
claims that a LUNMEB in $H^{d}\bigotimes H^{d}$ consists of at most
$d$ orthogonal vectors of non-maximally entangled states.

The unitary basic operators on $H^{d}$ used in [5] are given by
\begin{equation}\label{1}
U_{nm}=\sum_{k=0}^{d-1}e^{\frac{2\pi i}{d}nk}|k\oplus m\rangle\langle k|,
\end{equation}
where $i=\sqrt{-1}$.
Thus any unitary transformation $V$ can be represented as
\begin{equation}\label{2}
V=\sum_{p,q}f_{pq}U_{pq},
\end{equation}
where the coefficients $f_{pq}$ should satisfy some conditions such that $V$ is unitary.
Let $|\phi\rangle$ be a non-maximally entangled state in Schmidt form:
\begin{equation}\label{3}
|\phi\rangle=\sum_{k=0}^{d-1}C_{k}|kk\rangle,~~~~\sum_{k=0}^{d-1}C_{k}^2=1,
\end{equation}
where $C_{k}\geq0$, and the Schmidt rank of $|\phi\rangle$ is larger than one (at least two
non-zero Schmidt numbers $C_{k}$).
Start with $|\phi\rangle$ one can get $d$ states which are mutually orthogonal:
\begin{equation}\label{p0m}
    |\phi_{0m}\rangle=(U_{0m}\bigotimes I)|\phi\rangle,~~~~m=0,1,\cdots,d-1.
\end{equation}

To show that the set of states $|\phi_{0m}\rangle$ in (4) is unextendible, one has to show that
there does not exist unitary operator $V=\sum_{p,q}f_{pq}U_{pq}$
such that $|\Phi\rangle=(V\bigotimes I)|\phi\rangle$ is orthogonal
to all $|\phi_{0m}\rangle$. From
\begin{equation}\label{5}
    \langle\phi_{0m}|\Phi\rangle=\sum_{p=0}^{d-1}f_{pm}(C_{0}^{2}+e^{\frac{2\pi i}{d}p}C_{1}^{2}+
    \cdots+e^{\frac{2\pi i}{d}(d-1)p}C_{d-1}^{2})=0,~~~~\forall m,
\end{equation}
the authors in [5] derived that
\begin{equation}\label{cond}
\sum_{p=0}^{d-1}f_{pm}=0,~~~
\sum_{p=0}^{d-1}f_{pm}e^{\frac{2\pi i}{d}p}=0,~~~\cdots,~~~
\sum_{p=0}^{d-1}f_{pm}e^{\frac{2\pi i}{d}(d-1)p}=0,
\end{equation}
thereby $f_{pq}=0$, $\forall p,q$. However, this conclusion is not correct.
Since $\sum_{p=0}^{d-1}f_{pm}$,
$\sum_{p=0}^{d-1}f_{pm}e^{\frac{2\pi i}{d}p}$, $\cdots$, $\sum_{p=0}^{d-1}f_{pm}e^{\frac{2\pi i}{d}(d-1)p}$
are complex numbers, it is possible that some
$f_{pq}\neq 0$ while $\langle\phi_{0m}|\Phi\rangle=0,~\forall m$.
In the following, we give a simple counter example.

We consider the case of $d=4$. We take $C_0=C_2={1}/{\sqrt{3}}$, $C_1=C_3={1}/{\sqrt{6}}$ and begin with
the state
$$
|\phi\rangle=\frac{1}{\sqrt{3}}|00\rangle+\frac{1}{\sqrt{6}}|11\rangle+
\frac{1}{\sqrt{3}}|22\rangle+\frac{1}{\sqrt{6}}|33\rangle.
$$
The first four states which are mutually  orthogonal can be obtained
from (4) by using
the set of special unitary operators given in equation (\ref{1}):
\begin{eqnarray}\label{6}
  |\phi_{0m}\rangle&=&(U_{0m}\bigotimes I)\,|\phi\rangle \nonumber\\
   &=&\frac{1}{\sqrt{3}}|m\rangle|0\rangle+\frac{1}{\sqrt{6}}|1\oplus m\rangle|1\rangle+
\frac{1}{\sqrt{3}}|2\oplus m\rangle|2\rangle+\frac{1}{\sqrt{6}}|3\oplus m\rangle|3\rangle,
\end{eqnarray}
$m=0,1,2,3$.

To find a unitary operator $V=\sum_{p,q}f_{pq}U_{pq}$
such that $|\Phi\rangle=(V\bigotimes I)|\phi\rangle$ is orthogonal
to all vectors $|\phi_{0m}\rangle$, we take
\begin{equation}\label{fpq}
f_{10}=\frac{1-i}{2\sqrt{2}}, ~~~f_{30}=\frac{1+i}{2\sqrt{2}},~~~f_{11}=f_{31}=\frac{1}{2\sqrt{2}},~~~
f_{13}=-f_{33}=\frac{i}{2\sqrt{2}},
\end{equation}
and other $f_{pq}=0$. The unitary operator $V$ has a matrix form, under the ordered basis $|0\rangle$, $|1\rangle$, $|2\rangle$ and $|3\rangle$:
$$
V=\left(
\begin{array}{cccc}
\frac{1}{\sqrt{2}} & -\frac{1}{\sqrt{2}} &  & \\[2mm]
    \frac{1}{\sqrt{2}} & \frac{1}{\sqrt{2}} &  & \\[2mm]
     &  & -\frac{1}{\sqrt{2}} & \frac{1}{\sqrt{2}}\\[2mm]
     &  & -\frac{1}{\sqrt{2}} & -\frac{1}{\sqrt{2}}
\end{array}
\right).
$$
Therefore we have
\begin{equation}\label{9}
    \langle\phi_{0m}|\Phi\rangle=\frac{1}{3}\langle m|V|0\rangle+\frac{1}{6}\langle m\oplus1|V|1\rangle+
    \frac{1}{3}\langle m\oplus2|V|2\rangle+\frac{1}{6}\langle m\oplus3|V|3\rangle.
\end{equation}
It is easily checked that
$\langle\phi_{0m}|\Phi\rangle=0,~\forall\, m$. Hence we can get at least five
orthogonal vectors which are local unitary equivalent.
And the theorem in [5] does not hold in this case.

\setcounter{footnote}{0}
\renewcommand{\thefootnote}{\alph{footnote}}

\section{LUNMEB in $H^{2}\bigotimes H^{2}$}
\noindent
We have shown that the basis consisted of $d$ mutually orthogonal states (\ref{p0m})
is not unextendible generally.
In fact, to investigate thoroughly the existence and the numbers of
LUNMEB existed for a given system, it is not enough to use the
particular unitary operators (\ref{1}) in constructing the basis.
One should use general unitary operators $V$ to find the second
vector $|\Phi\rangle=(V\bigotimes I)|\phi\rangle$
that is orthogonal to the given non-maximally entangled state $|\phi\rangle$,
and repeat the procedure to find the basic vectors, until the
condition (iv) in the first section applies. However, for general
$H^{d}\bigotimes H^{d}$ system it is formidable
to solve such problems completely. In the following, we
give a complete solution to the problem of locally unextendible non-maximally entangled
basis for the case of $H^{2}\bigotimes H^{2}$.

We start with a non-maximally entangled state in the Schmidt decomposition form:
\begin{equation}
|\phi\rangle=C_{0}|00\rangle+C_{1}|11\rangle,
\end{equation}
where $C_{0}\neq C_{1}$ are non-zero and $C_{0}^2+C_{1}^2=1$.
The number of states in a LUNMEB is relevant to the choice of the
second vector. Let the second vector be
$|\Phi\rangle=(V\bigotimes I)|\phi\rangle$, with $V$ an arbitrary  unitary operator such that,
up to a global phase factor,
$$
V\,(|0\rangle,|1\rangle)=(|0\rangle,|1\rangle)
\left(
\begin{array}{cc}
\cos\theta & -\sin\theta\, e^{i\theta_{1}}\\[2mm]
\sin\theta\, e^{i\theta_{2}}& \cos\theta\, e^{i\theta_{3}}
\end{array}
\right).
$$
From $\langle\phi|\Phi\rangle=C_{0}^{2}\cos\theta+C_{1}^{2}\cos\theta\, e^{i\theta_{3}}=0$, we have
$\cos\theta=0$, and $V|0\rangle=e^{i\alpha}|1\rangle$, $V|1\rangle=e^{i\beta}|0\rangle$, since
if $\cos\theta\neq0$, then
$e^{i\theta_{3}}=-{C_{0}^{2}}/{C_{1}^{2}}$ is a real number, which
gives rise to $C_{0}=C_{1}$, and leads a contradiction.

For the possible construction of the third vector, suppose
$|\Psi\rangle=(U\bigotimes I)|\phi\rangle$ is
orthogonal to both $|\phi\rangle$ and $|\Phi\rangle$. From the similar
reason to the construction of $|\Phi\rangle$, we have
$U|0\rangle=e^{i\mu}|1\rangle$, and $U|1\rangle=e^{i\eta}|0\rangle$.
From
\begin{eqnarray}
\langle\Psi|\Phi\rangle &=& C_{0}^{2}\langle0|U^{\dagger}V|0\rangle+C_{1}^{2}\langle1|U^{\dagger}V|1\rangle\nonumber\\[3mm]
   &=&  C_{0}^{2}e^{i(\alpha-\mu)}+C_{1}^{2}e^{i(\beta-\eta)}=0,\nonumber
\end{eqnarray}
we obtain $e^{i(\alpha-\mu-\beta+\eta)}=-{C_{1}^{2}}/{C_{0}^{2}}$,
which implies $C_{0}=C_{1}$. Therefore the third basis $|\Psi\rangle$ does not exist, and
$|\phi\rangle$ and $|\Phi\rangle$ form a LUNMEB.

\section{Conclusion and discussion}
\noindent
The unextendible product basis (UPB),
the unextendible maximally entangled basis (UMEB) and the
locally unextendible non-maximally entangled basis (LUNMEB)
in $H^{d}\bigotimes H^{d}$ are of significance in the theory of
quantum information. However, the results obtained so far are far from being satisfied.
Even for UMEB, one can only solve the problem completely for the case of $d=2$ \cite{4}.

We have studied the locally unextendible non-maximally entangled basis.
In correcting an error in \cite{5}, we have shown that
there could be more than $d$ orthogonal vectors in a LUNMEB for $H^{d}\bigotimes
H^{d}$ systems. Like the UMEB, we completely solved the problem of LUNMEB for the case of $d=2$.
The approach we used can be applied to the study of LUNMEB for high dimensional cases.
Nevertheless, the problem becomes complicated as $d$ increases.

\nonumsection{References}


\begin{thebibliography}{000}
\bibitem{1}
C. H. Bennett, D. P. Divincenzo, T. Mor, P. W. Shor, J. A. Smolin
and B. M. Terhal (1999), {\it Unextendible product bases and bound
entanglement}, Phys. Rev. Lett. 82, 5385.

\bibitem{2}
D. P. Divincenzo, T. Mor, P. W. Shor, J. A. Smolin and B. M. Terhal
(2003), {\it Unextendible product bases, uncompletable  product
bases and bound  entanglement}, Commun. Math. Phys. 238, pp.
379-410.

\bibitem{3}
R. Duan, {\it Super-activation of zero-error capacity of noise quantum channels}, arXiv:0906.2527.

\bibitem{4}
S. Bravyi, J. A. Smolin, {\it Unextendible maximally entangled bases}, arXiv:0911.4090v2.

\bibitem{5}
I. Chakrabarty, P. Agrawal and A. K. Paty (2012), {\it Locally unextendible non-maximally entangled basis}, Quant. Inf. Comput. 12, 0271.

\end{thebibliography}
\end{document}